# RF Power Transport for Accelerators

*Eric Montesinos*
CERN, Geneva, Switzerland

**Abstract**
This paper reviews the main types of radio-frequency power transport systems which may be used in accelerators. It gives essentials on rectangular waveguides and coaxial lines. Basics of combining systems, splitting systems and transmission lines are discussed.

## 1 Introduction

The purpose of this article is to describe the transmission of RF power of several kW up to several MW at frequencies from the MHz to GHz range. The RF power generated by an RF generator must be transported and distributed to a load or cavity or several loads or cavities.

The main requirements of the transmission lines are to have low loss, low reflections, high reliability, adjustment of phase and amplitude, …

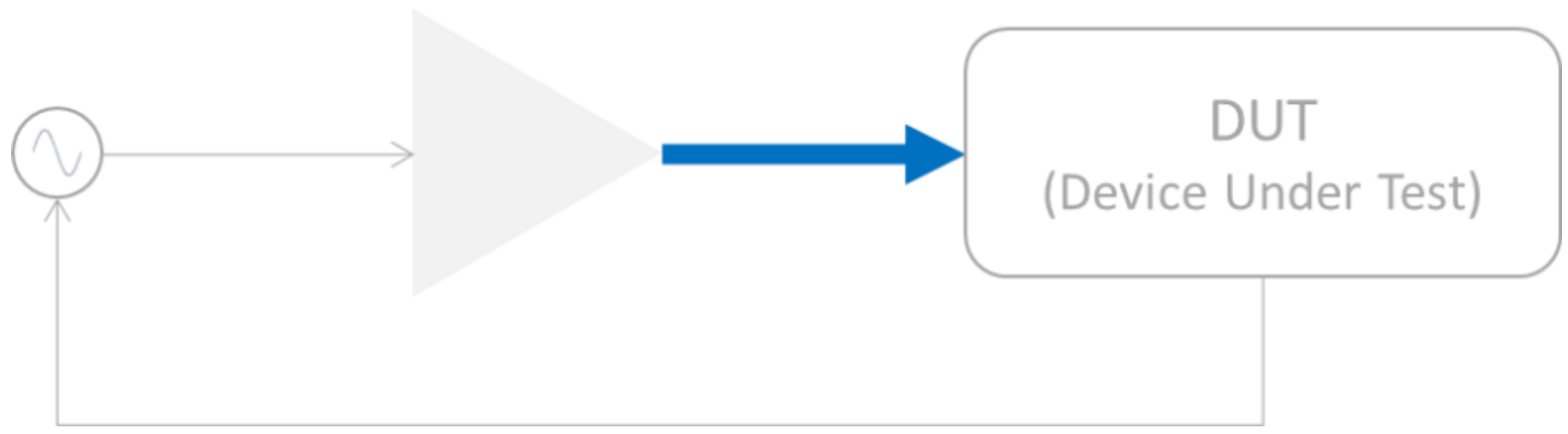

**Fig. 1:** A very simple representation showing that the connection between the amplifier and the Device Under Test (DUT) that in accelerators can be cavity or kicker, is done with a transmission line.

## 2 RF power transmission lines families

Once we have the required RF output power, it has to be transported from the RF amplifier output to the load. Several transmission lines exist, however, the main lines that are used in high power RF are the rectangular waveguides and the coaxial lines.

### 2.1 Two wires lines

These lines are often used for indoor antenna, radio or TV, but as radiating to the environment, cannot be used for high power Transportation.

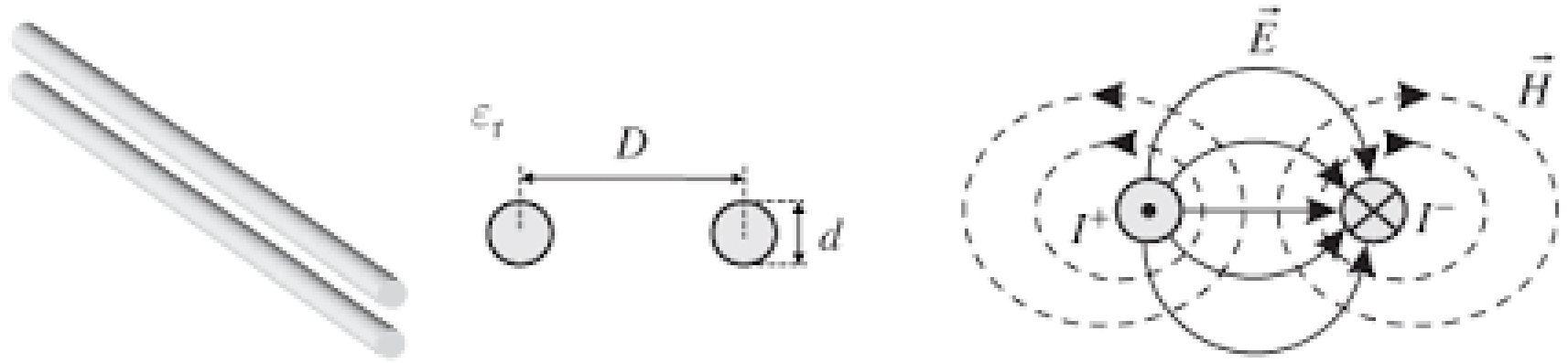

**Fig. 2:** A two wires transmission line is radiating to the environment.

## 2.2 Strip-lines

The strip-lines are often used for microwave integrated circuits, however as they radiate to the environment and as they have limited power capability, they cannot be used for high power transportation.

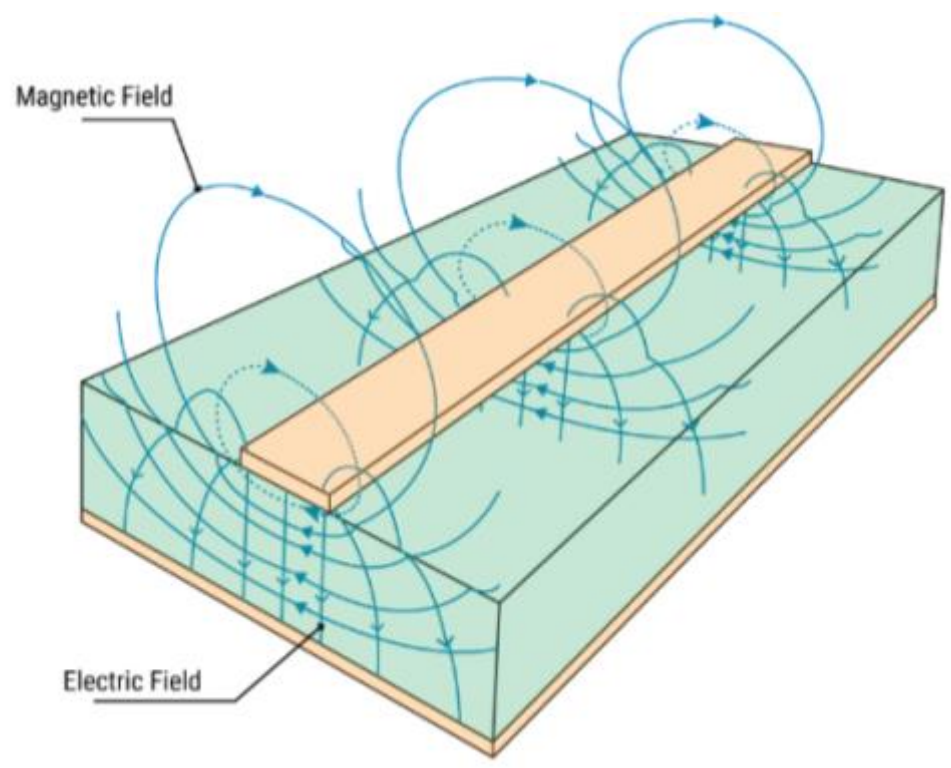

**Fig. 3:** A strip-line transmission line is radiating to the environment.

## 2.3 Coaxial lines

The coaxial lines are often used for power RF transmission and connection of RF components. However, due to their high loss above a certain frequency, due to heating of inner conductor and dielectric material, they can only be used over a limited frequency range and with some power limitations that we will describe later on.

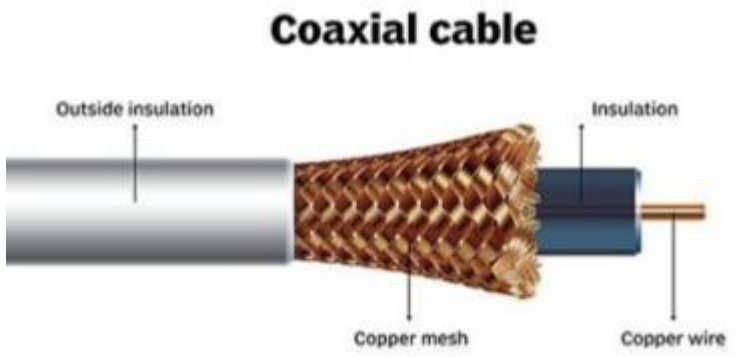

**Fig. 4:** A cable is well suited to transmit a modest amount of power.

## 2.4 Waveguides (rectangular, cylindrical, elliptical)

Waveguide is the solution often used for high power RF transmission. It is mostly the rectangular waveguide that is used as normalized, even if it requires quite some waveguide plumbing and imposes quite some rigidity.

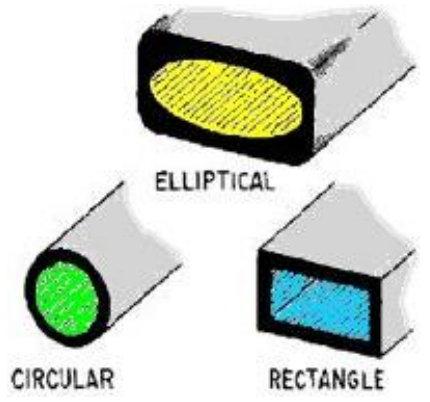

**Fig. 5:** A waveguide is a good solution to transmit up to an important amount of power.

## 3 Skin depth effect

RF only needs a few µm of good electrical layer to flow along:

- at the surface, conductivity is 100%
- at one skin depth, it is decreased to 36.8%
- at two skin depths, 13.5%
- at three skin depths, 5.0%
- at four, 1.8%
- at five, 0.7%.

When d = 5 x δ, more than 99 % of the current flows in the conductor.

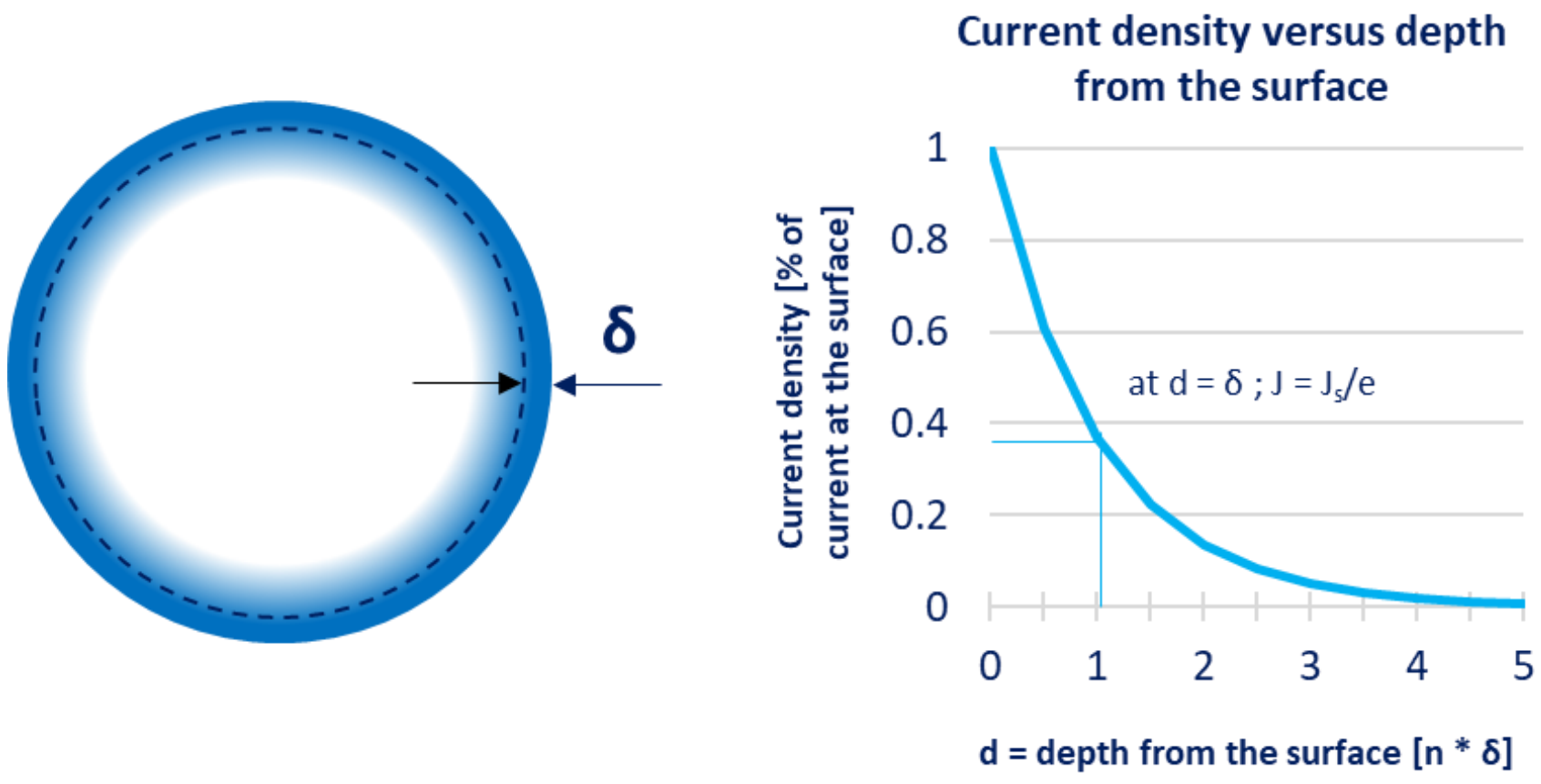

**Fig. 6:** RF current is flowing through only the first layers of a plain cable.

$$J = Js\, e^{-\left(\frac{d}{\delta}\right)} \tag{1}$$

$$\delta = \sqrt{\frac{2\rho}{\omega\mu}} \tag{2}$$

with:

J = current density

Js = current density at the surface

d = depth from the surface

δ = skin depth in which 63 % of the current flows

ρ = resistivity of the conductor

ω = 2πf,

μ = μr * μ0

μr = relative magnetic permeability of the conductor

μ0 = permeability of free space

For copper at 400 MHz, ρ = 1.678 * 10-8 Ωm, μr = 0.999991, δ = 3.26 μm.

Another way to write Eq. (2) allows to better understand the effect of the material is:

$$\boldsymbol{\delta} = \sqrt{\frac{\boldsymbol{\rho}}{\boldsymbol{\pi f \mu}}} \tag{3}$$

with:

ρ = resistivity of the conductor

f = frequency

$\mu = \mu_r * \mu_0$

$\mu_r$ = relative magnetic permeability of the conductor

$\mu_0$ = permeability of free space.

Doing so, one can summarize in the following table, for some conventional material, the thickness needed to transmit RF power adequately.

| material | P [nΩm] | $\mu_r$ | δ @ 200 MHz [μm] | 5 x δ @ 200 MHz [μm] |
|---|---|---|---|---|
| Gold | 24.4 | 1 | 5.56 | 27.8 |
| Silver | 15.9 | 1 | 4.49 | 22.5 |
| Copper | 17.2 | 1 | 4.67 | 23.4 |
| Aluminium | 28.2 | 1 | 5.97 | 29.9 |
| Tin | 109 | 1 | 11.75 | 58.8 |
| Lead | 220 | 1 | 16.70 | 83.5 |

**Fig. 7:** Thickness of material needed to allow an adequate RF transmission.

## 4 Rectangular waveguides (WG)

The main advantage of the waveguides is that waveguides provide propagation with low loss.

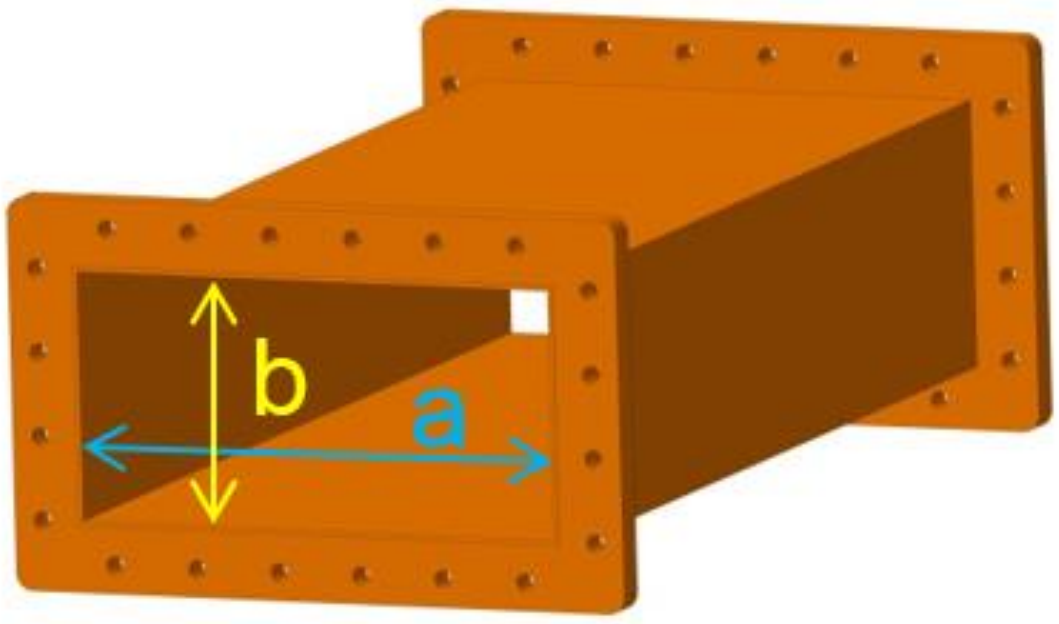

**Fig. 8:** Example of a rectangular waveguide. The size ***a*** is the width, and the size ***b*** is the height.

The Main parameters of a rectangular waveguide are given by the following equations:

Waveguide wavelength $$\boldsymbol{\lambda g} = \frac{\lambda}{\sqrt{1-\left(\frac{\lambda}{2a}\right)^2}} \quad (4)$$

Cut-off frequency dominant mode $$\boldsymbol{fc} = \frac{c}{2a} \quad (5)$$

Cut-off frequency next higher mode $$\boldsymbol{fc2} = \frac{c}{4a} \quad (6)$$

Usable frequency range **1.3 fc to 0.9 fc2**. (7)

From the transmission line Eqs. (4)-(7), we can see that the waveguides are usable only over certain frequency ranges. For very lower frequencies the waveguide dimensions become impractically large. For very high frequencies the waveguide dimensions become impractically small, and the manufacturing tolerance becomes a significant portion of the waveguide size. Figure 9 lists some of the currently used waveguide sizes. The EIA standard names the waveguides with respect to their width. So, a WR2300 waveguide has a width of 23.00 inches. This is very convenient to quickly identify the size and the reference of the waveguides. It is also common to have half height waveguide, when the RF power is not too high. Indeed, as it can been seen within Eqs. (4)-(6), the height is not providing any limitation.

| Waveguide name | | | Recommended frequency band of operation (GHz) | Cutoff frequency of lowest order mode (GHz) | Cutoff frequency of next mode (GHz) | Inner dimensions of waveguide opening (inch) | Inner dimensions of waveguide opening (mm) |
|---|---|---|---|---|---|---|---|
| EIA | RCSC | IEC | | | | | |
| WR2300 | WG0.0 | R3 | 0.32 — 0.45 | 0.257 | 0.513 | 23.000 × 11.500 | 584.2 x 292.1 |
| WR1150 | WG3 | R8 | 0.63 — 0.97 | 0.513 | 1.026 | 11.500 × 5.750 | 292.1 x 146 |
| WR340 | WG9A | R26 | 2.20 — 3.30 | 1.736 | 3.471 | 3.400 × 1.700 | 86.1 x 43.2 |
| WR75 | WG17 | R120 | 10.00 — 15.00 | 7.869 | 15.737 | 0.750 × 0.375 | 19.05 x 9.52 |
| WR10 | WG27 | R900 | 75.00 — 110.00 | 59.015 | 118.03 | 0.100 × 0.050 | 2.54 x 1.27 |
| WR3 | WG32 | R2600 | 220.00 — 330.00 | 173.571 | 347.143 | 0.0340 × 0.0170 | 0.86 x 0.43 |

**Fig. 9:** Some of the commonly used standard waveguides.

### 4.1 Rectangular waveguides peak power

The peak power limitation for such waveguides is given by the following equation:

$$P = 6.63\ 10^{-4}\ Emax^2 \sqrt{b^2(a^2 - \frac{\lambda^2}{4})} \tag{8}$$

with:

P = Power in watts

a = width of waveguide in cm

b = height of waveguide in cm

λ = free space wavelength in cm

Emax = breakdown voltage gradient of the dielectric filling the waveguide in Volt/cm (for dry air 30 kV/cm, for ambient air 10 kV/cm).

Looking at this equation, we can notice that each waveguide size will have its own characteristic. Figure 10 illustrates the peak power limitation per waveguide size. However, and this is a very important point, the values provided in Figure 10 are theoretical values for straight waveguides, in practise, one should consider ***the limit being 10% of this value***.

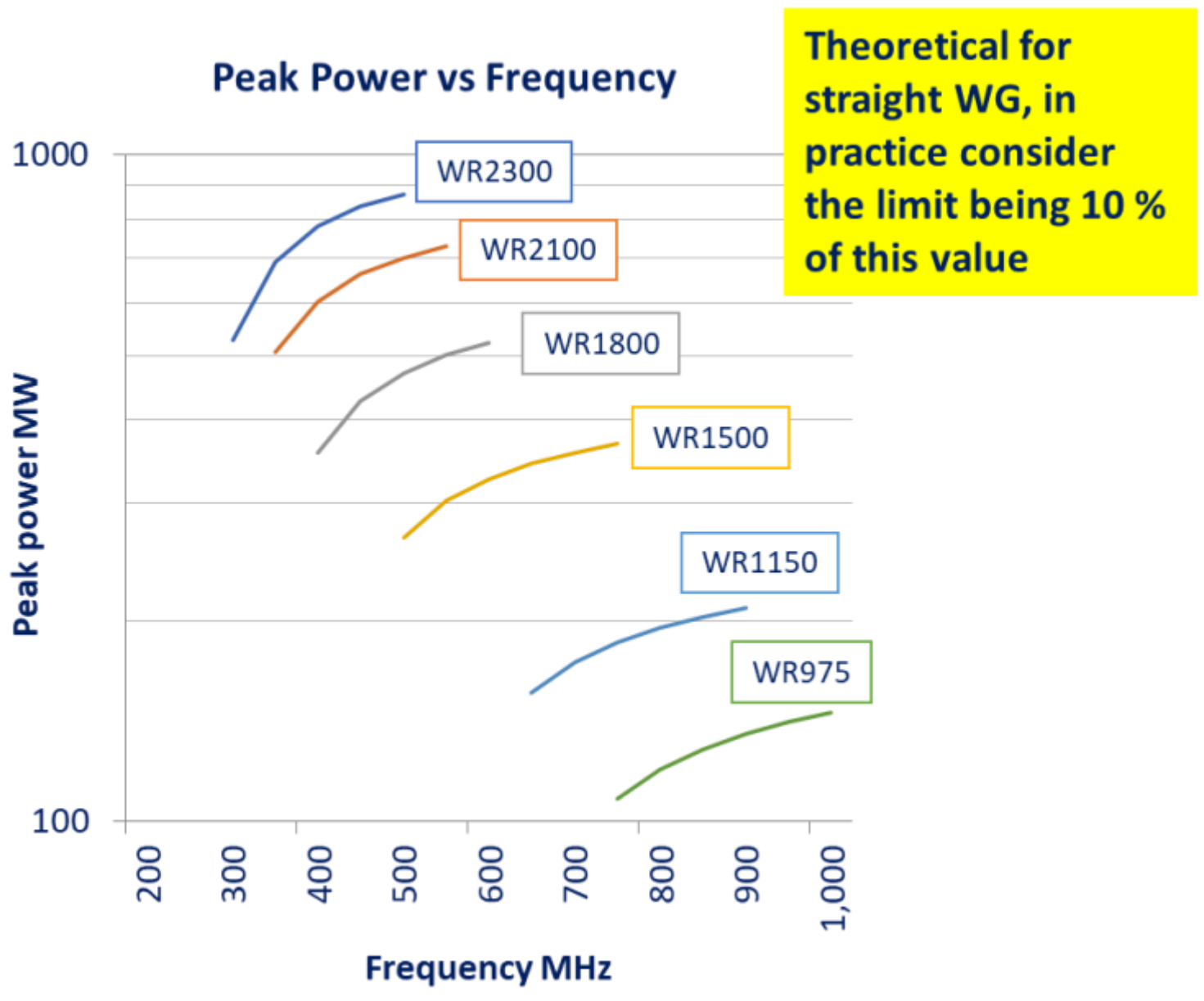

**Fig. 10:** RF frequency range and peak power per waveguide size.

## 4.2 Rectangular waveguides attenuation

The walls of the waveguides are not perfect conductors, they have finite conductivity resulting in losses. The attenuation of the line is also dependant of the geometry:

$$\alpha = \frac{4a0}{a}\frac{\sqrt{c/\lambda}}{\sqrt{1-(\lambda/2a)^2}}\left(\frac{a}{2b}+\frac{\lambda^2}{4a^2}\right) \quad (9)$$

with:

α = value of attenuation in dB/m

a0 = 3 10-7 [dB/m] for copper

a = width of waveguide in m

b = height of waveguide in m

λ = free space wavelength in m.

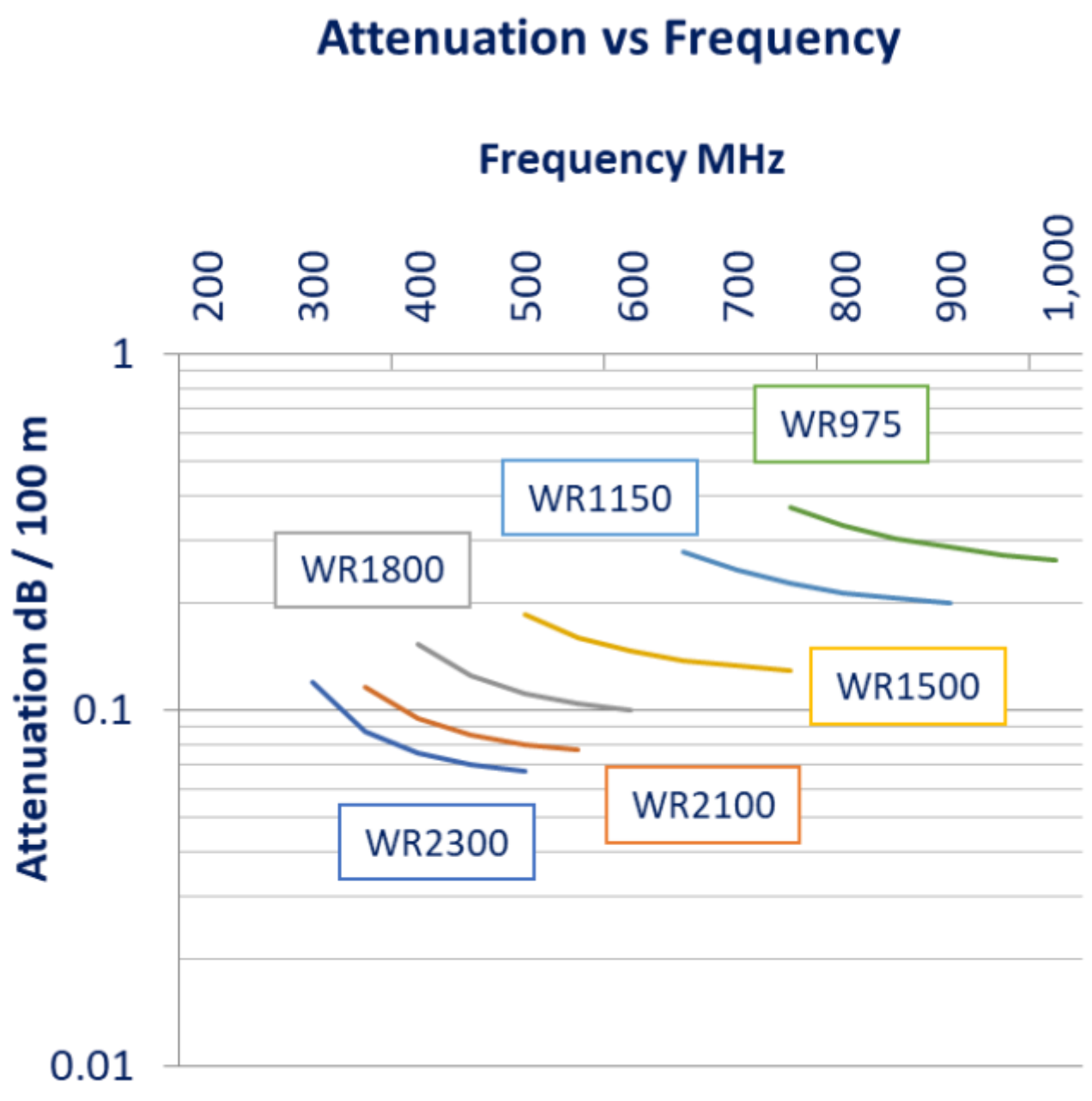

**Fig. 11:** Attenuation of a copper waveguide full height size.

| Attenuation factors of waveguides made from different material normalized to a waveguide of same size made of copper | |
|---|---|
| Copper | 1.00 |
| Silver | 0.98 |
| Aluminium | 1.30 |
| Brass | 2.05 |

**Fig. 12:** Table showing how material can affect the attenuation of a rectangular waveguide.

## 5 Coaxial lines

As we have seen, for the lower frequencies, the waveguide dimensions become impractically large. The Coaxial lines are then one of the most commonly used solution.

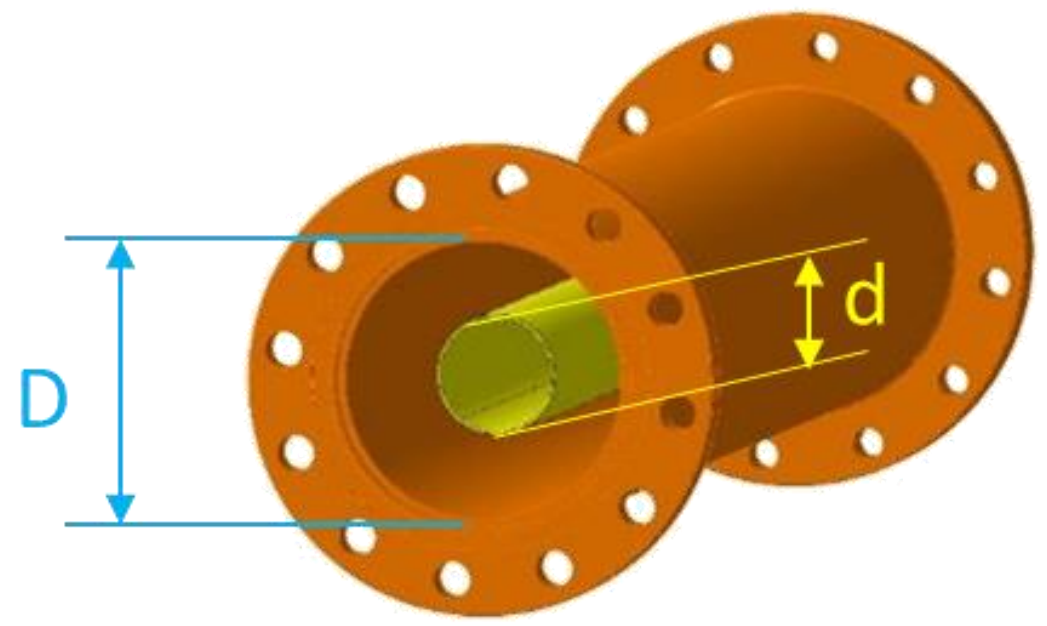

**Fig. 13:** Example of a coaxial line. The size ***d*** is the outer diameter of the inner line, and the size ***D*** is the inner diameter of the outer line.

The characteristic impedance of a coaxial line is given by the following equation:

$$\mathbf{Zc} = \frac{\mathbf{60}}{\sqrt{\varepsilon r}} \mathbf{ln}\left(\frac{D}{d}\right) \tag{10}$$

with

D = inner dimension of the outer conductor

d = outer dimension of the inner conductor

$\varepsilon_r$ = dielectric characteristic of the medium.

The dielectric characteristic of the medium plays a very important role in a coaxial line, and it has not to be neglected. Indeed, coaxial cables are often with PTFE foam to keep concentricity, flexible lines have spacer helicoidally placed all along the line, and rigid lines are made of two rigid tubes maintained concentric with supports. Regarding the spacers used, the size of the inner and outer diameters will have to be compensated.

| | Outer conductor | | Inner conductor | |
|---|---|---|---|---|
| **Size** | **Outer diameter** | **Inner diameter** | **Outer diameter** | **Inner diameter** |
| 7/8" | 22.2 mm | 20 mm | 8.7 mm | 7.4 mm |
| 1 5/8" | 41.3 mm | 38.8 mm | 16.9 mm | 15.0 mm |
| 3 1/8" | 79.4 mm | 76.9 mm | 33.4 mm | 31.3 mm |
| 4 1/2" | 106 mm | 103 mm | 44.8 mm | 42.8 mm |
| 6 1/8" | 155.6 mm | 151.9 mm | 66.0 mm | 64.0 mm |

**Fig. 14:** Some of the commonly used standard coaxial lines.

## 5.1 Coaxial lines peak power

The Power handling of an air coaxial line is related to breakdown field E. The peak power limitation for such coaxial lines is given by the following equations:

$$\boldsymbol{Vpeakmax = E\frac{d}{2}ln\left(\frac{D}{d}\right)} \tag{11}$$

$$\boldsymbol{Ppeakmax = \frac{Vpeakmax^2}{2Zc}} \tag{12}$$

$$\boldsymbol{Ppeakmax = \frac{E^2\ d^2\sqrt{\varepsilon r}}{480}ln\left(\frac{D}{d}\right)} \tag{13}$$

with:

E = breakdown strength of air (‘dry air’ E = 3 kV/mm, commonly used value is E = 1 kV/mm for ambient air)

D = inside electrical diameter of outer conductor in mm

d = outside electrical diameter of inner conductor in mm

Zc= characteristic impedance in Ω

εr = relative permittivity of dielectric

f = frequency in MHz.

Looking at these equations, we can notice that the size of the coaxial line strongly impacts its characteristic. Figure 15 illustrates the peak power limitation for some normalized coaxial line sizes. However, and this is a very important point, the values provided in Fig. 15 are theoretical values for straight coaxial line, in practise, one should consider the limit to be reduced for other pieces of lines, a reduction of -30% should be applied to an elbow for example.

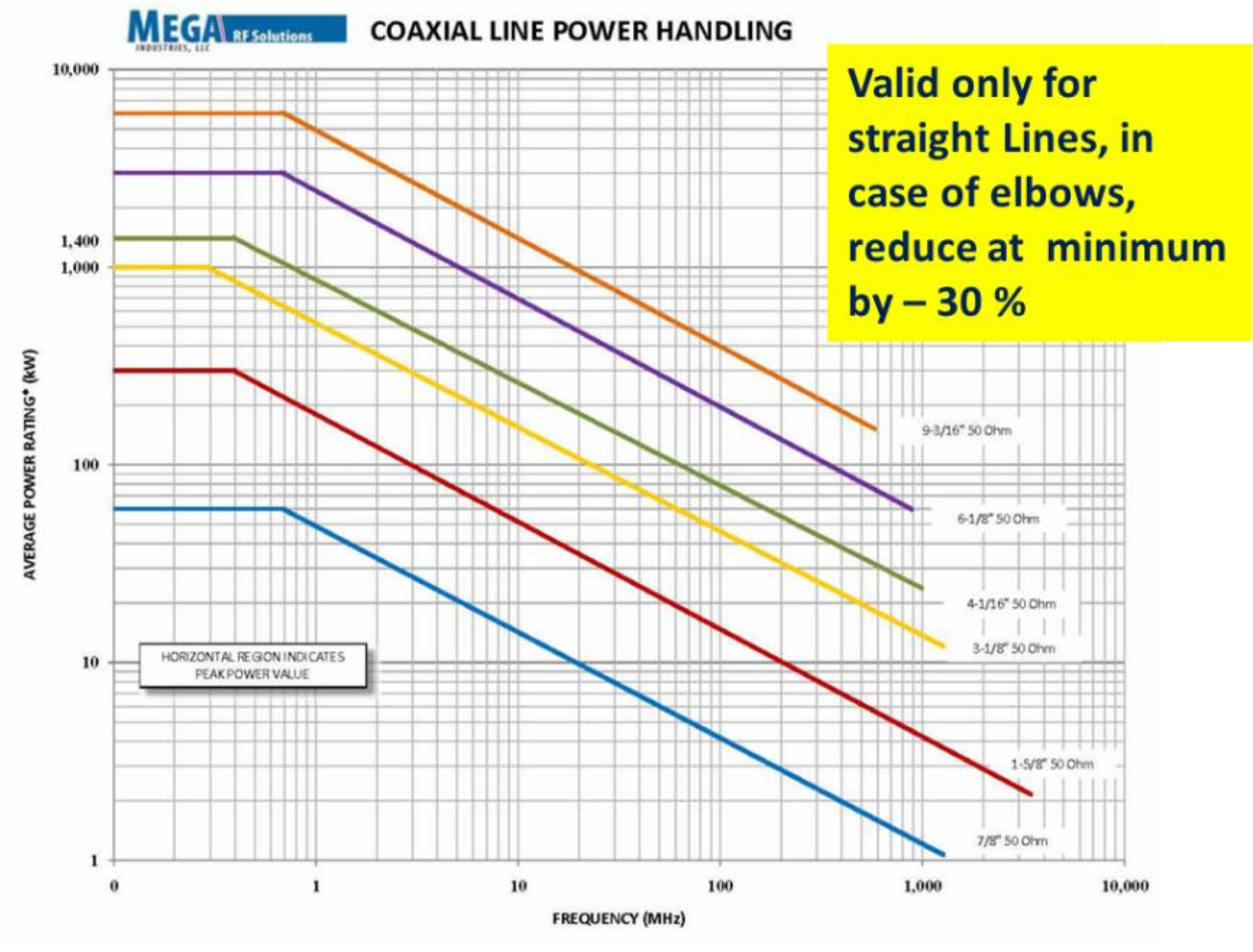

**Fig. 15:** RF frequency range and peak power per coaxial line size.

## 5.2 Coaxial lines attenuation

The The attenuation of a coaxial line can be approximated with the following expression:

$$\boldsymbol{\alpha} = \left(\frac{\mathbf{36.1}}{\boldsymbol{Zc}}\right)\left(\frac{\mathbf{1}}{\boldsymbol{D}} + \frac{\mathbf{1}}{\boldsymbol{d}}\right)\sqrt{\boldsymbol{f}} + \mathbf{9.1}\ \sqrt{\boldsymbol{\varepsilon r}}\ \boldsymbol{tan\delta}\ \boldsymbol{f}, \tag{14}$$

where:

α = attenuation constant in dB/m

Zc= characteristic impedance in Ω

f = frequency in MHz

D = inside electrical diameter of outer conductor in mm

d = outside electrical diameter of inner conductor in mm

εr = relative permittivity of dielectric tan δ = loss factor of dielectric.

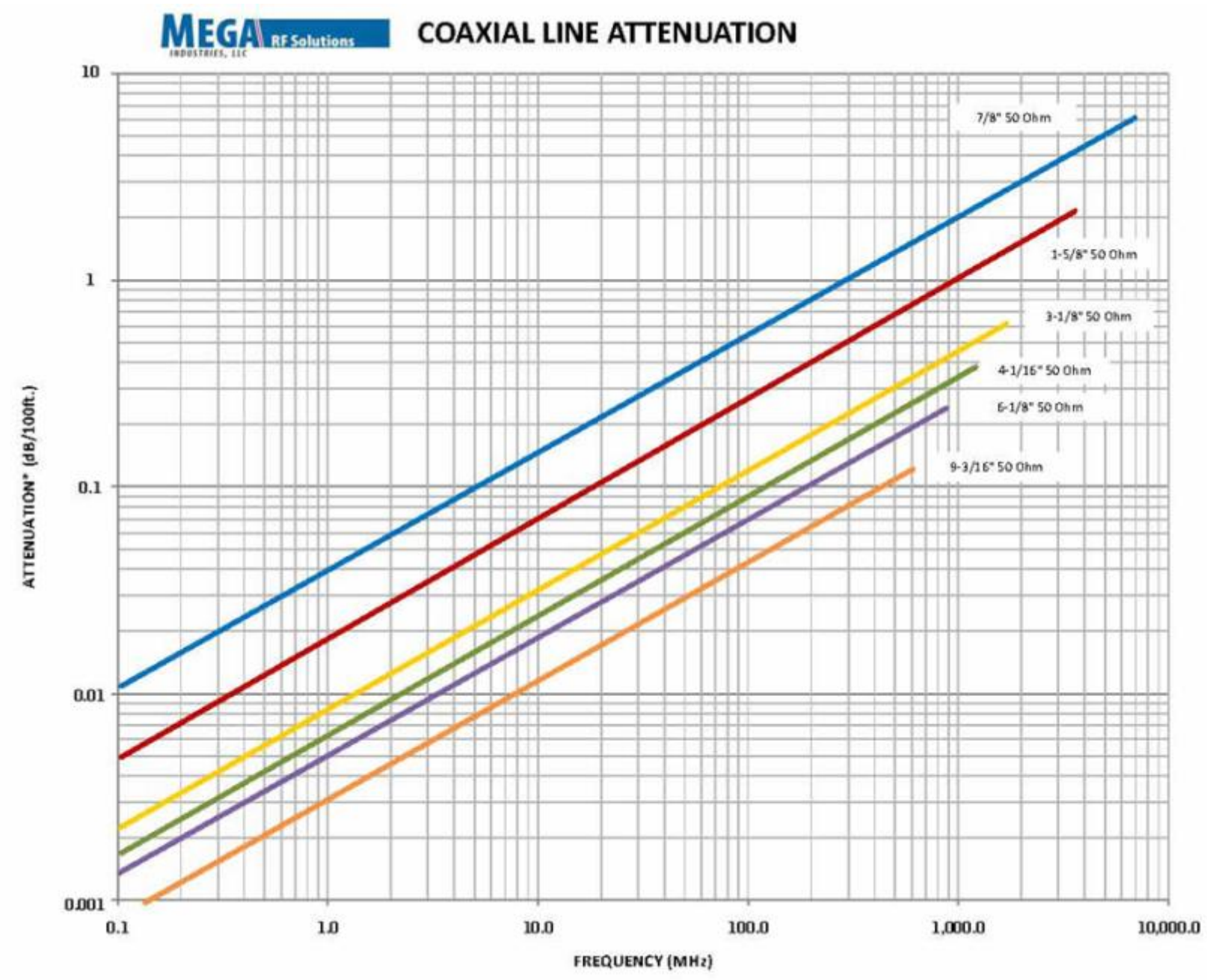

**Fig. 16:** Attenuation of some standardized coaxial lines.

When selecting the coaxial line, the best compromise has to be made with respect to the needs of the project between: size, peak power capability, and attenuation.

| Material | $\varepsilon_r$ | tan δ | Breakdown MV/m |
|---|---|---|---|
| Air | 1.00006 | 0 | 3 |
| Alumina 99.5% | 9.5 | 0.00033 | 12 |
| PTFE | 2.1 | 0.00028 | 100 |

**Fig. 17:** The medium will impact the attenuation of the line.

It is always very important to take all the parameters into consideration and to remind that the power limitations given by the suppliers must carefully and strictly be followed. A damaged coaxial line by mechanical deformation or by overheating will change its impedance characteristic, and this will considerably reduce its power handling capability.

### 5.3 Coaxial lines impedance

Taking all Eqs. (10)-(13) together, one can plot a graph summarizing losses versus impedance with respect to power, voltage and attenuation.

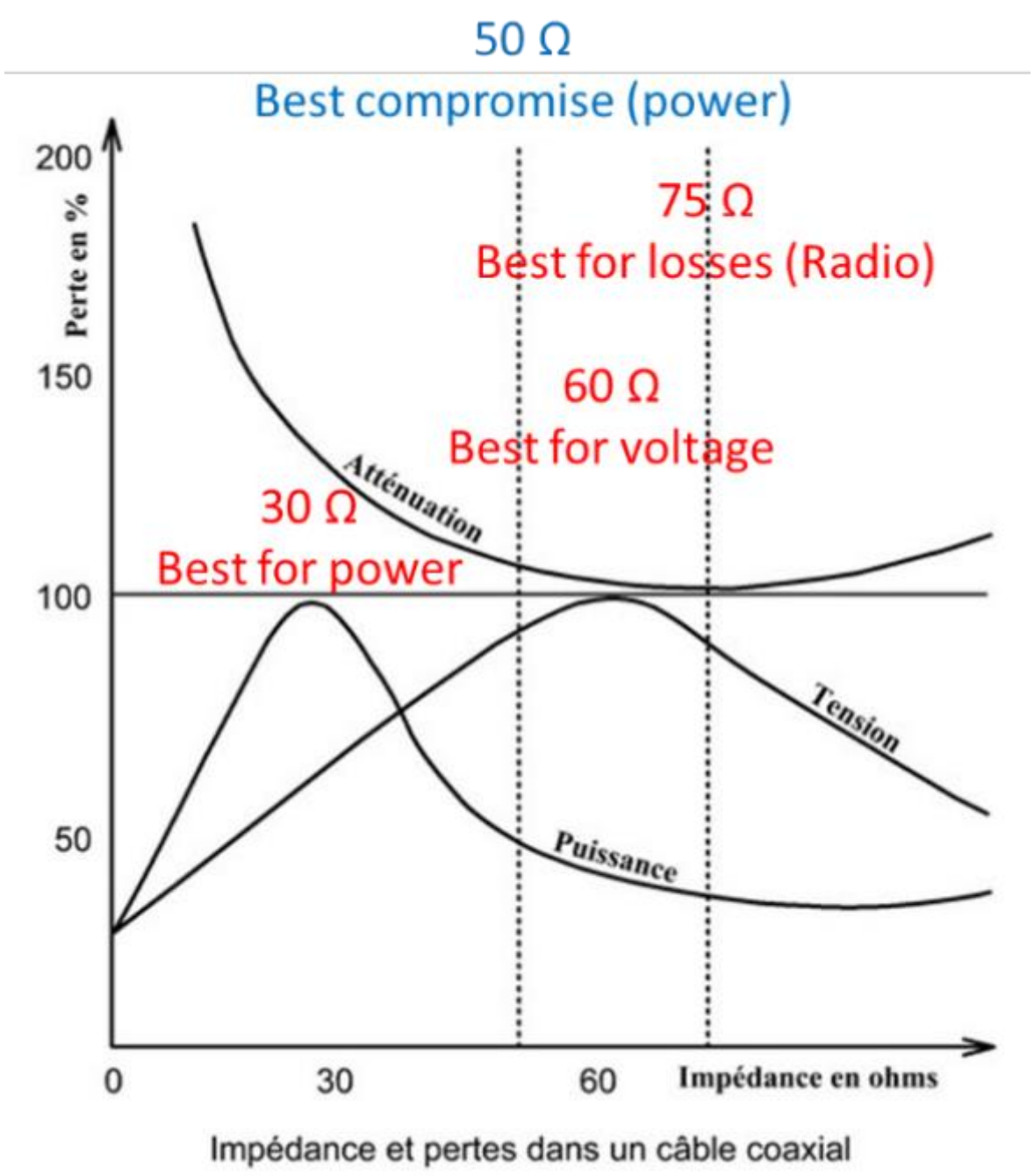

**Fig. 18:** There is not a best impedance, only a compromise to be made regarding your project.

A compromise to normalize line construction and instrumentation was chosen at 50 Ω.

## 6 Reflection from load and circulator

A major phenomenon that has to be consider whilst selecting the correct transmission line is the matching of the impedance.

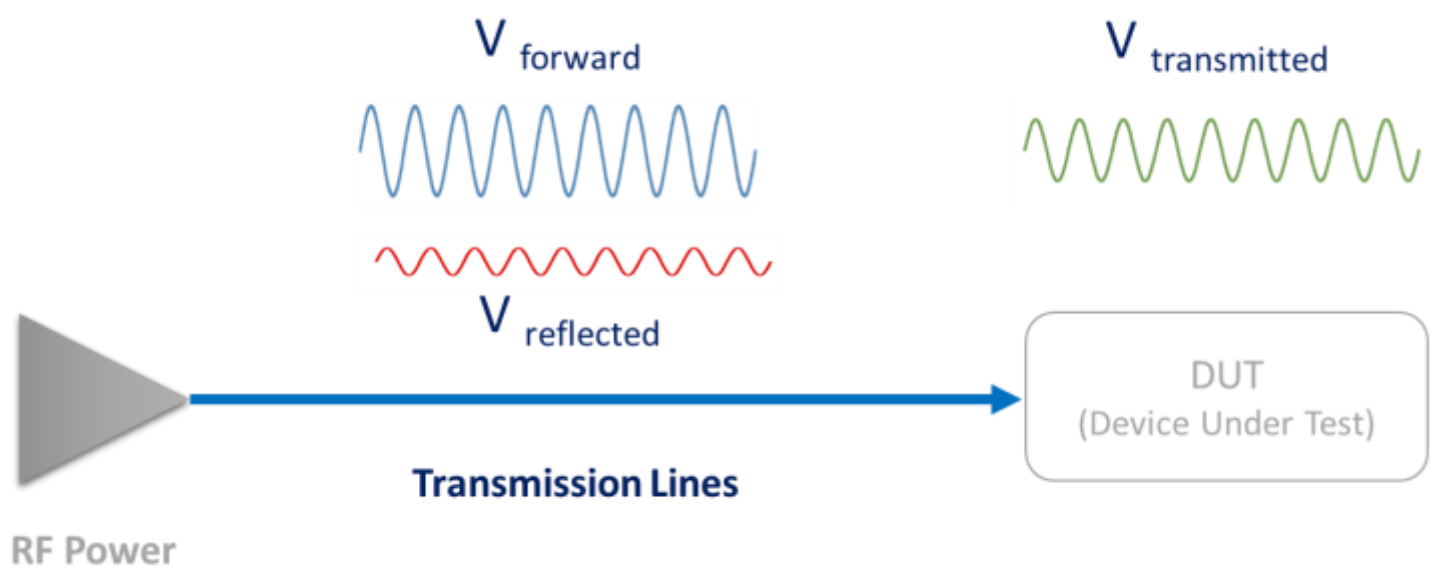

**Fig. 19:** Values used to define the mismatch of your system.

The Standing Wave Ration (SWR) is a measure of impedance matching of the Device Under Test (DUT). A wave is partly reflected when a transmission line is terminated with other than a pure resistance equal to its characteristic impedance.

The reflection coefficient is defined by:

$$\Gamma = \frac{\mathbf{Vr}}{\mathbf{Vf}} \tag{15}$$

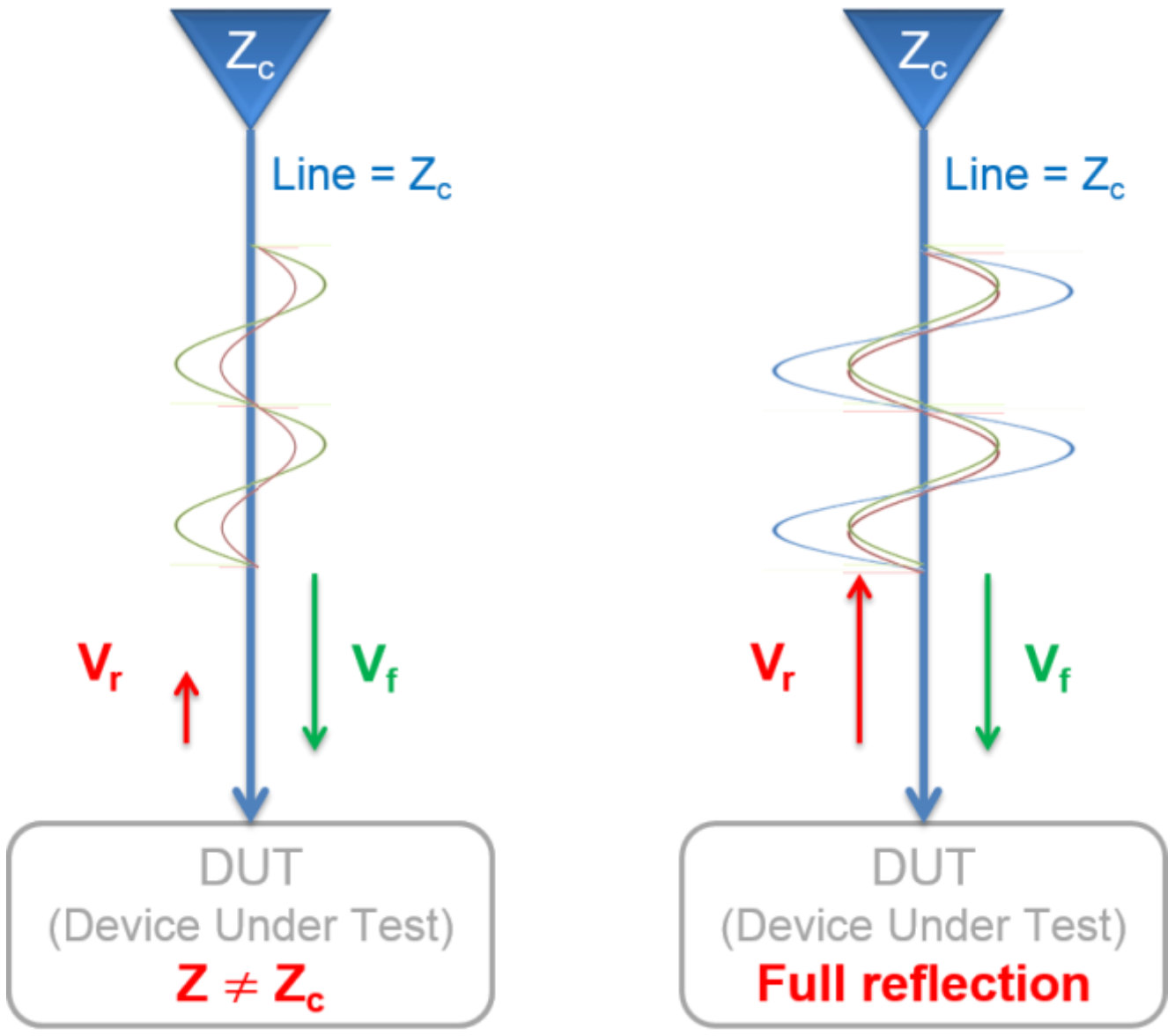

**Fig. 20:** Forward and Reflected waves regarding the impedance of the DUT.

| | |
|---|---|
| $\Gamma$ = -1 | when the line is short-circuited<br>complete negative reflection |
| $\Gamma$ = 0 | when the line is perfectly matched, no reflection |
| $\Gamma$ = 1 | when the line is open-circuited<br>complete positive reflection |

**Fig. 21:** SWR regarding the impedance of the DUT.

At some points along the line the forward and reflected waves are exactly in phase, and then:

$$|\mathbf{Vmax}| = |\mathbf{Vf}| + |\mathbf{Vr}| = |\mathbf{Vf}| + |\Gamma\mathbf{Vf}| = (\mathbf{1} + |\Gamma|)\ |\mathbf{Vf}| \tag{16}$$

In case of full reflection full reflection:

$$|\mathbf{Vmax}| = \mathbf{2}\ |\mathbf{Vf}| \tag{17}$$

At other points the forward and reflected waves are 180° out of phase, and then:

$$|\mathbf{Vmin}| = |\mathbf{Vf}| - |\mathbf{Vr}| = |\mathbf{Vf}| - |\Gamma\mathbf{Vf}| = (\mathbf{1} - |\Gamma|)\,|\mathbf{Vf}| \tag{18}$$

In case of full reflection full reflection:

$$|\boldsymbol{Vmin}| = \mathbf{0} \tag{19}$$

The Voltage Standing Wave Ratio (VSWR) is defined by:

$$\mathbf{VSWR} = \frac{|\mathbf{Vmax}|}{|\mathbf{Vmin}|} = \frac{\mathbf{1}+|\Gamma|}{\mathbf{1}-|\Gamma|}\,. \tag{20}$$

So, we have seen Eq. (17) that in case of full reflection $V_{max} = 2\ V_f$, this means that $P_{max}$ is equivalent to 4 $P_f$. Here we have to be careful as RF power people are often stating the maximum power 'is' four time in the transmission line in case of full reflection. This is somehow an abuse of language, as it is only the voltage that is double, and the current that is doubled, but both are not in phase under these full reflection conditions. We must remain very careful and clearly state that maximum power 'is equivalent' to four time the forward power. An explanation is that all the datasheets from the suppliers are given in power, not in voltage. So, if we want to operate a system that has to sustain full reflection, we have to select the correct line from the supplier taking into account four times the forward power.

In any cases, RF power amplifiers will not like this reflected wave. Klystron output cavity is disturbed and Grid tube, IOT and Transistor voltage capability have to be fully respected in order not to damage the devices. A swift protection if $P_r > P_{rmax}$ has to be implemented. Unfortunately, this solution, as illustrated in Fig. 41, makes the system not operational, which is not always acceptable.

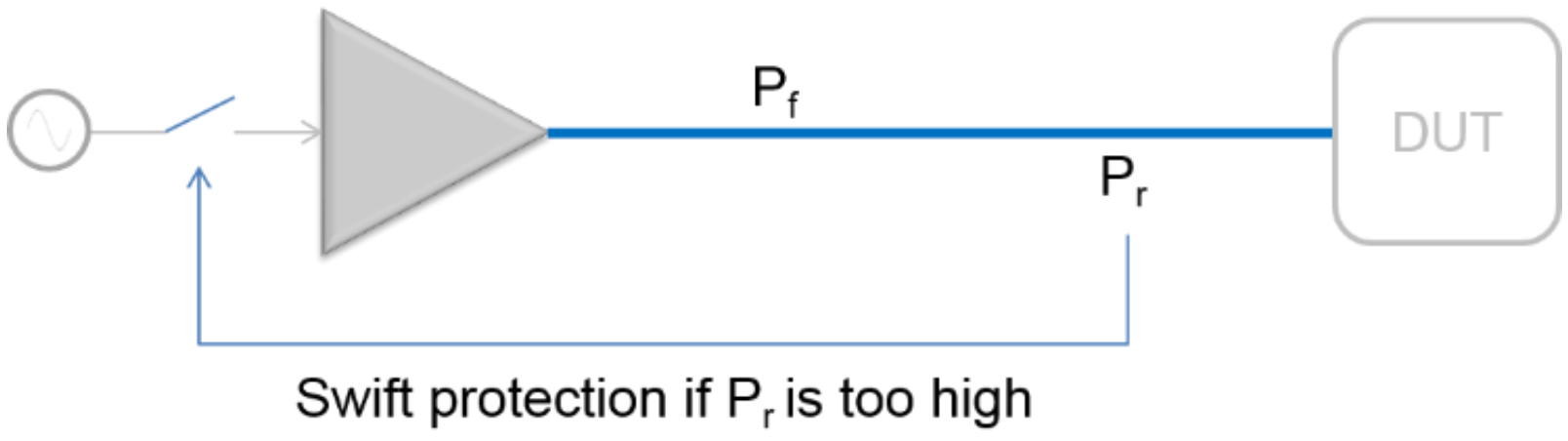

**Fig. 22:** Swift protection to protect the DUT.

In order to protect our lines and our amplifiers from this reflected power a specific component, the circulator, has been developed. This is a passive non-reciprocal three ports device. As shown in figure 23, the signal entering any port is transmitted only to the next port in rotation. The best place to insert it is close to the reflection source. The lines between circulator and DUT shall sustain four times $P_f$ in case of full reflection. A load of $P_f$ is needed on port 3 to absorb $P_r$.

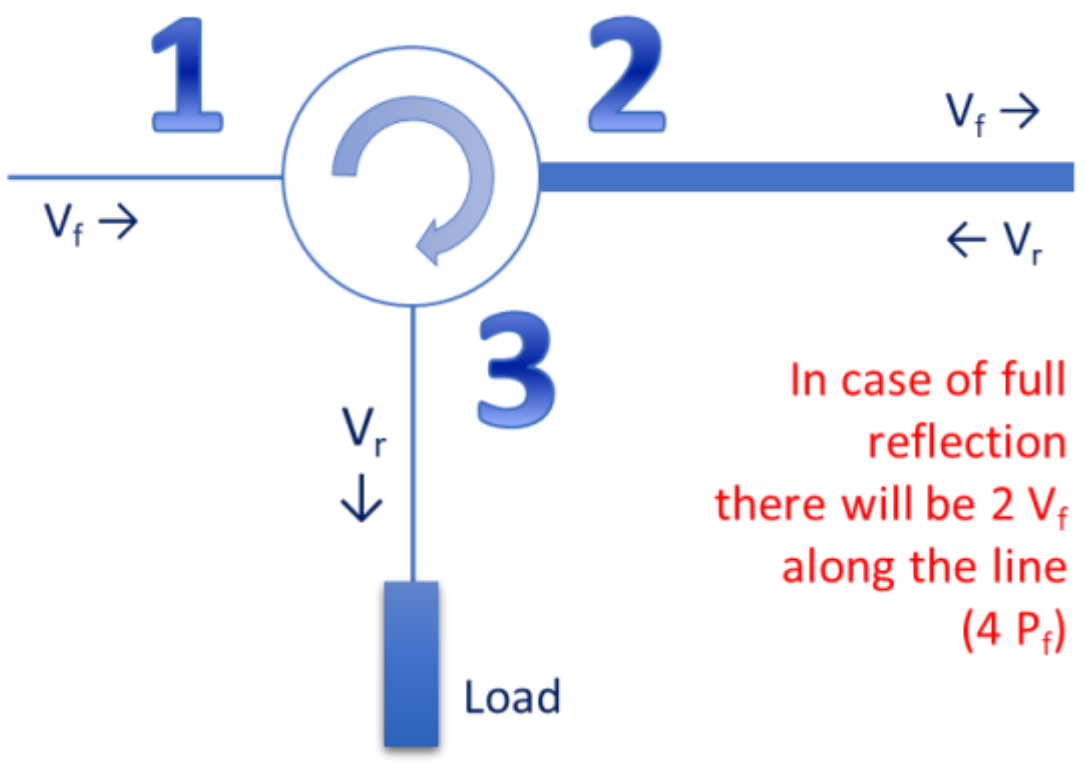

**Fig. 23:** Circulator basic principle. A device to correctly protect the DUT.

Even in case of full reflection $V_{max}$ = 2 $V_f$ and so $P_{max}$ equivalent to 4 $P_f$ the RF power amplifiers will not see the reflected power and will not be affected. The lines between circulator and DUT must at least be designed for 4 $P_f$ and the load must be designed for $P_f$. The main advantage of such a configuration shown in Fig. 24, is that the system remains always operational.

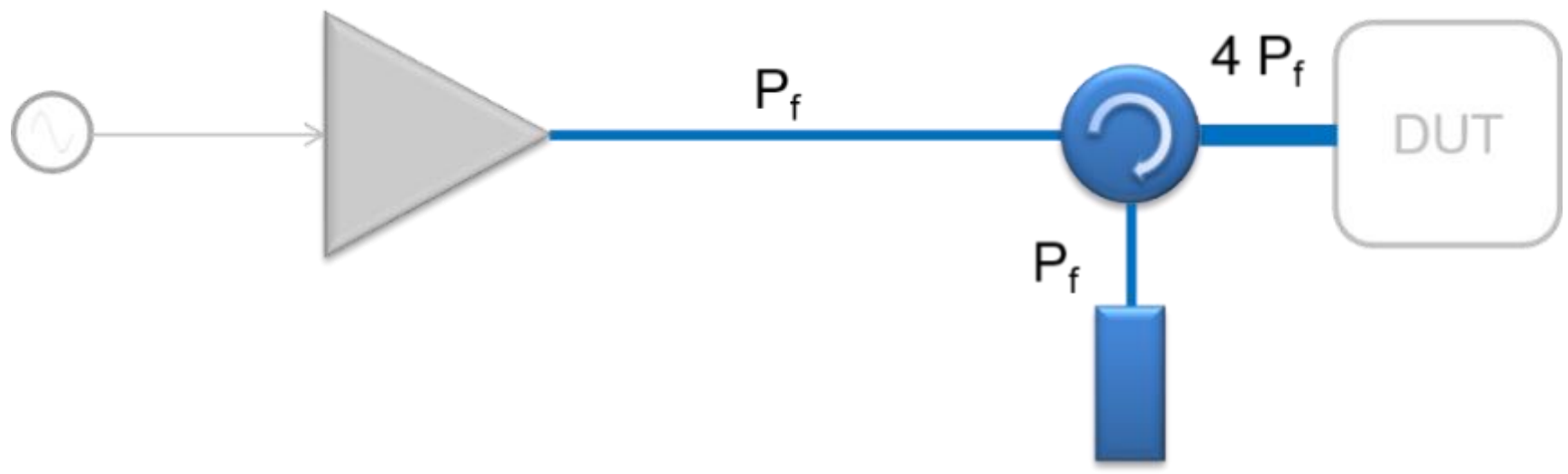

**Fig. 24:** A system protected with a circulator always remains operational.

The most misunderstood concept of circulators is that of isolation. Circulators do not provide isolation until one of the ports is terminated. Then the isolation between the other two ports (in the direction opposing the direction of circulation) is approximately equal to the return loss due to any mismatch on the terminated port. So, a very good load is needed on port 3 in order to guaranty a good isolation at port 1.

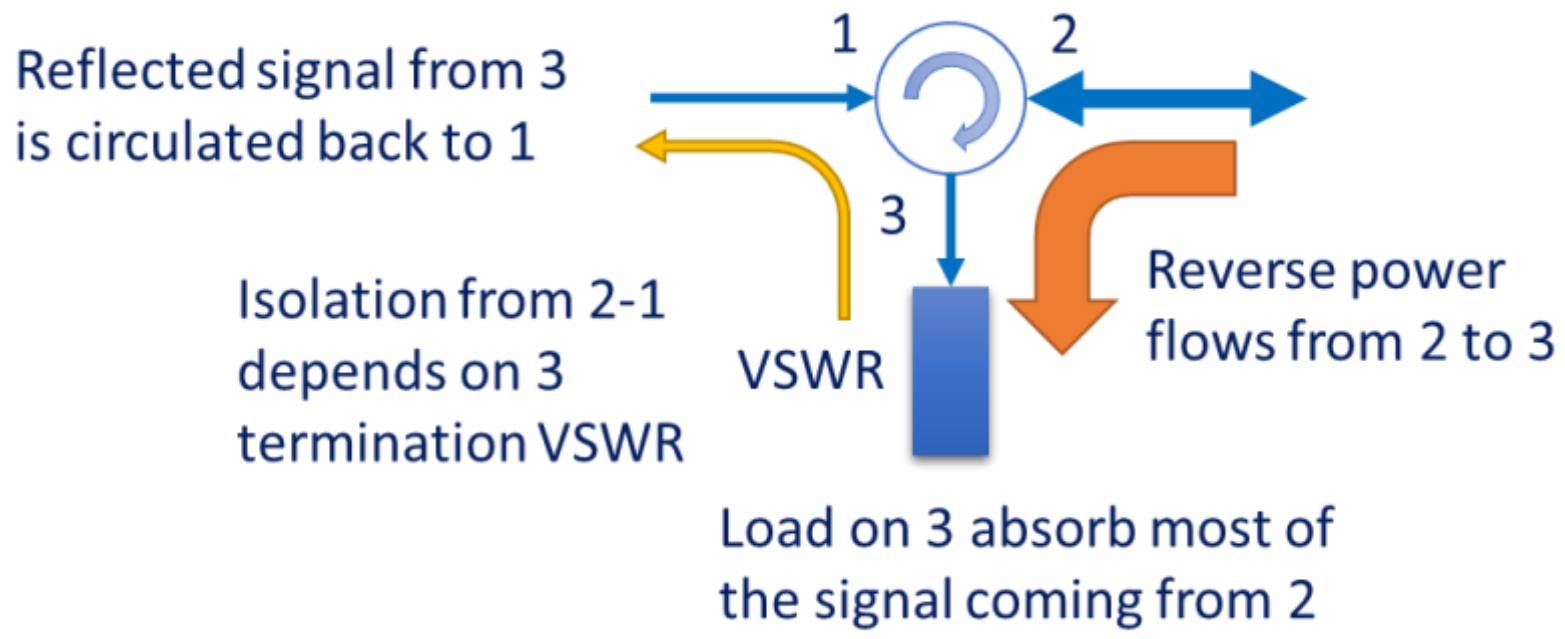

**Fig. 25:** A good load on port 3 is mandatory to ensure a correct isolation.

# 7 Fundamental Power Coupler (FPC)

To complete the description of the transmission power chain, there is a final device that ensure the transfer of power from the line to the DUT, the Fundamental Power Coupler (FPC). The FPC is the connecting part between the RF transmission line and the RF cavity. It is a specific piece of transmission line that also must provide the vacuum barrier for the beam vacuum. FPC are one of the most critical parts of the RF cavity system in an accelerator. A good RF design, a good mechanical design and a high-quality fabrication are essential for an efficient and reliable operation. Illustrated in figure 26 are some of the FPC recently designed at CERN in the past few years.

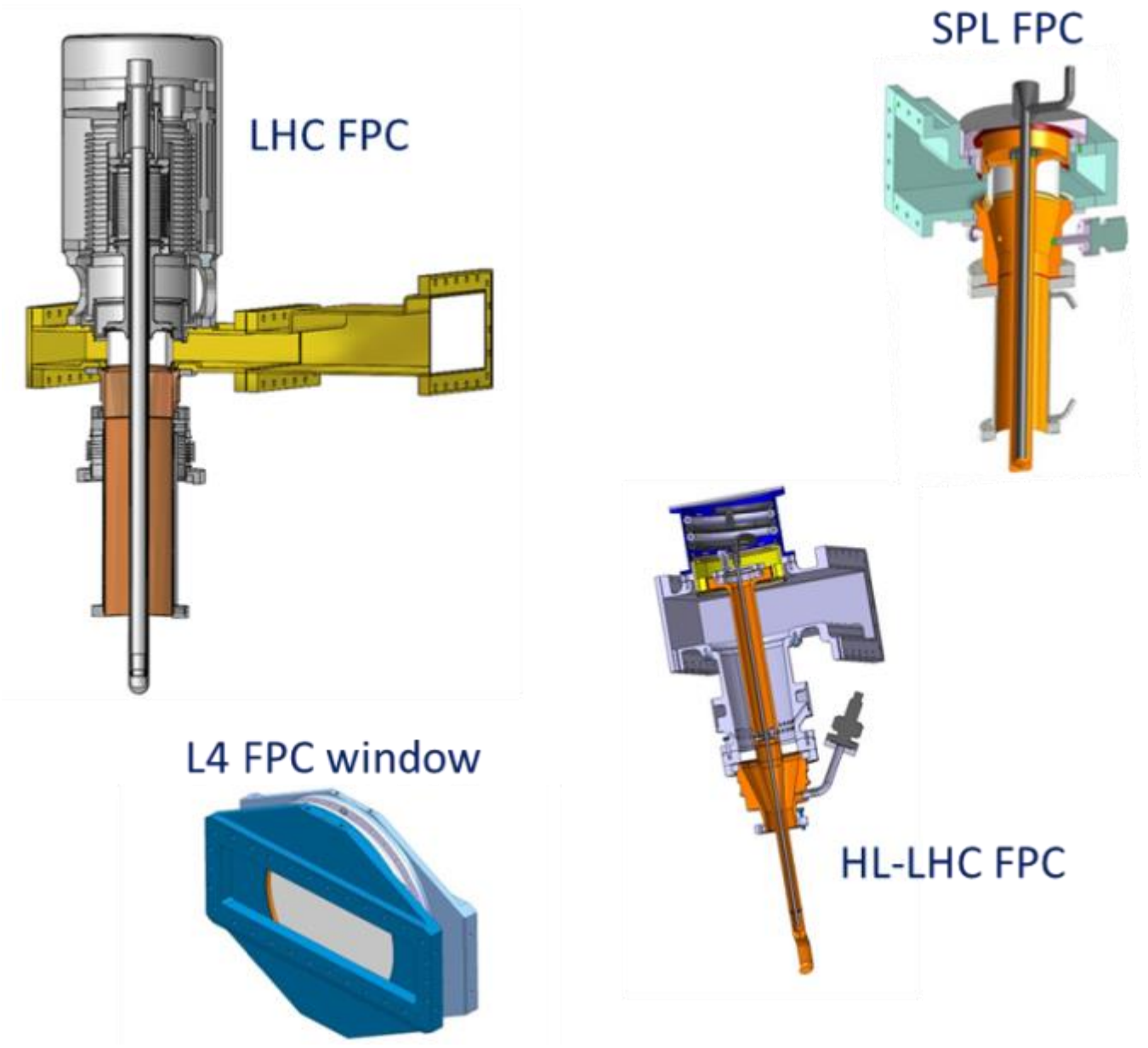

**Fig. 26:** Various CERN FPC developed in the past few years.

Over the last decades, CERN designed quite some couplers that allowed to establish a power rating given the frequency, the list is detailed in Fig. 27.

| | |
|---|---|
| **LHC** | **400 MHz, 500 kW CW SW** |
| **SPS 2.0** | **200 MHz, 750 kW CW TW** |
| **SPL 2.0** | **704 MHz, 900 kWp 10 % SW** |
| **SPL 3.0** | **704 MHZ, 1000 kWp 10 % SW** |
| **Linac4** | **352 MHz, 1000 kWp 10 % SW** |
| **Crab DQW** | **400 MHz, 100 kW CW SW** |
| **Crab RFD** | **400 MHz, 100 kW CW SW** |
| **ESRF** | **352 MHz, 200 kW CW SW** |
| **SOLEIL** | **352 MHz, 200 kW CW SW** |
| **APS 1.0** | **352 MHz, 200 kW CW SW** |
| **SPS LIU** | **200 MHz, 800 kW CW TW** |
| *HG (SPL 3.0)* | *704 MHz, 1500 kWp 10 % SW* |
| *LHC 2.0* | *400 MHz, 500 kW CW SW* |
| *APS 2.0* | *352 MHz, 250 kW CW SW* |

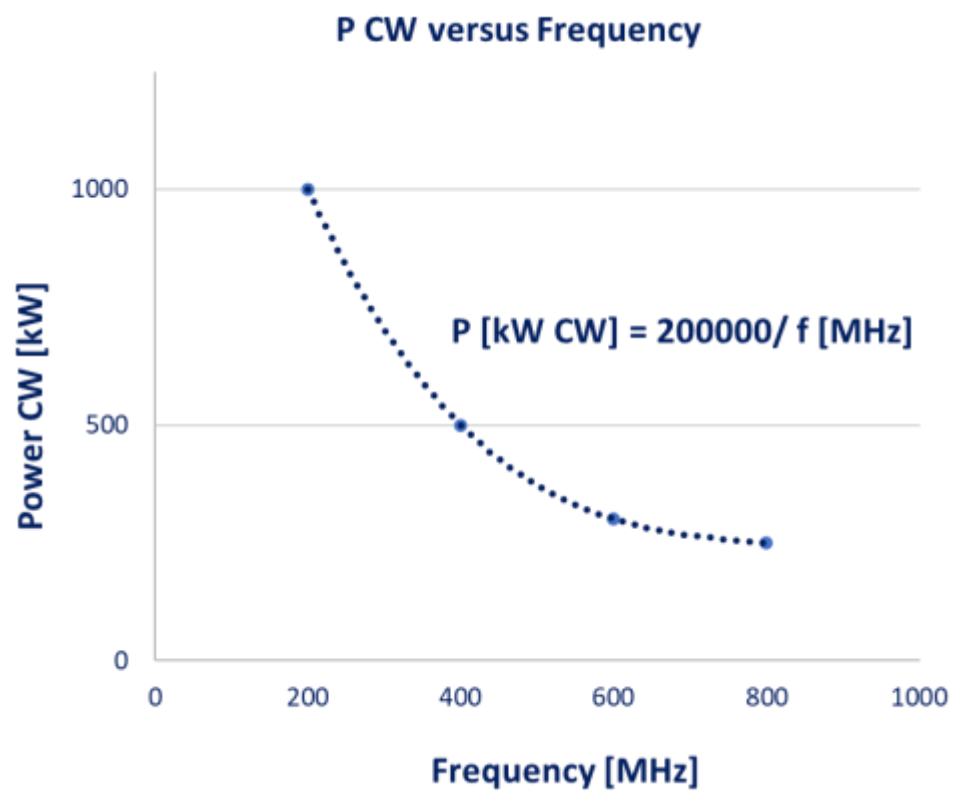

**Fig. 27:** Relation between Continuous wave (CW) power of CERN's developed FPC versus frequency.

Even if not a strict limit (we aim to double the power capability for the Future Circular Collider (FCC) at CERN), one can notice that with the current technologies involved in the coupler construction, the maximum power ratings is given by the following equation:

$$\textbf{Power [kW CW]} = \frac{\textbf{200000}}{\textbf{frequency [MHz]}} \ . \tag{21}$$

## 7.1 Ceramic window

A special item needs a few lines on it, the ceramic window. This is the most important device of a FPC. It ensures the vacuum leak tightness of the FPC, and of the entire machine! Any leak on the window immediately leads into degradation of the cavity and of the machine. It is commonly a ceramic brazed with metal.

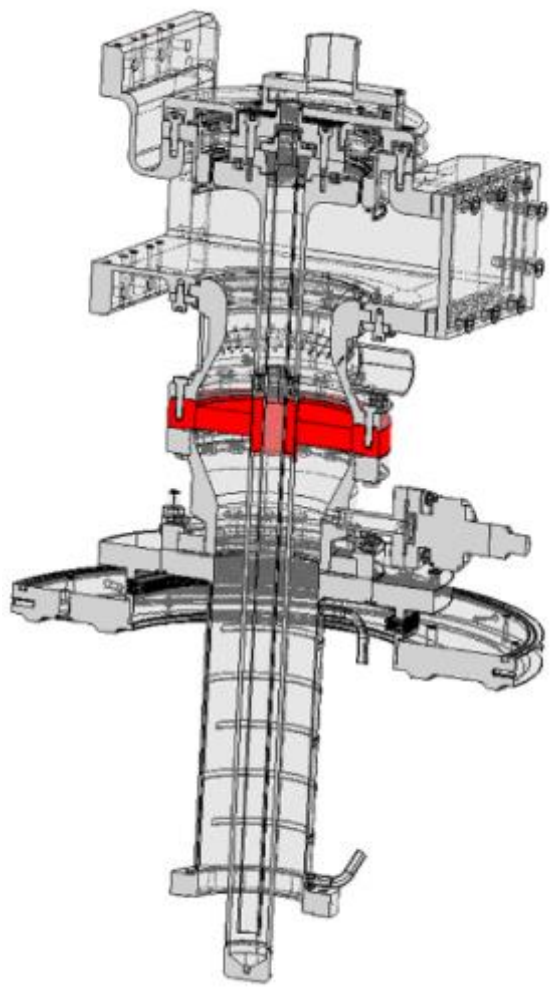

**Fig. 28:** The ceramic window is a key item in a power coupler design.

## 7.2 Ceramic window material

Most of the windows are built with an Al2O3 ceramic. A very important parameter is the purity of the ceramic. A too pure ceramic will be with very few losses, that is perfect for RF power, but will be very difficult to braze as the metallization will not adhere on it. A ceramic with impurities will be much easier to braze but will have a lot of losses that will induce a difficult cooling.

| | Purity | RF losses | Brazing |
|---|---|---|---|
| $Al_2O_3$ | 99.9 % | Very Low | Very difficult |
| $Al_2O_3$ | 97.6 % | Medium | Medium |
| $Al_2O_3$ | 95 % | Higher | Easier |

**Fig. 29:** Summary of the advantages and difficulties of ceramic with respect to their purity.

In view of future machines, we push R&D to move to 99.5 % purity, having less losses, allowing for more power.

### 7.3 Ceramic window metallisation

Before brazing the metallic line, the window has to be metallized. The most common medium used is a Molly-Manganese deposition on the surfaces to be brazed. It is often painted by hands. This paint is very sensible and must be kept in movement at any time, under a controlled temperature and humidity. The metallic lines will be brazed onto that MoMn support, it is of the highest importance.

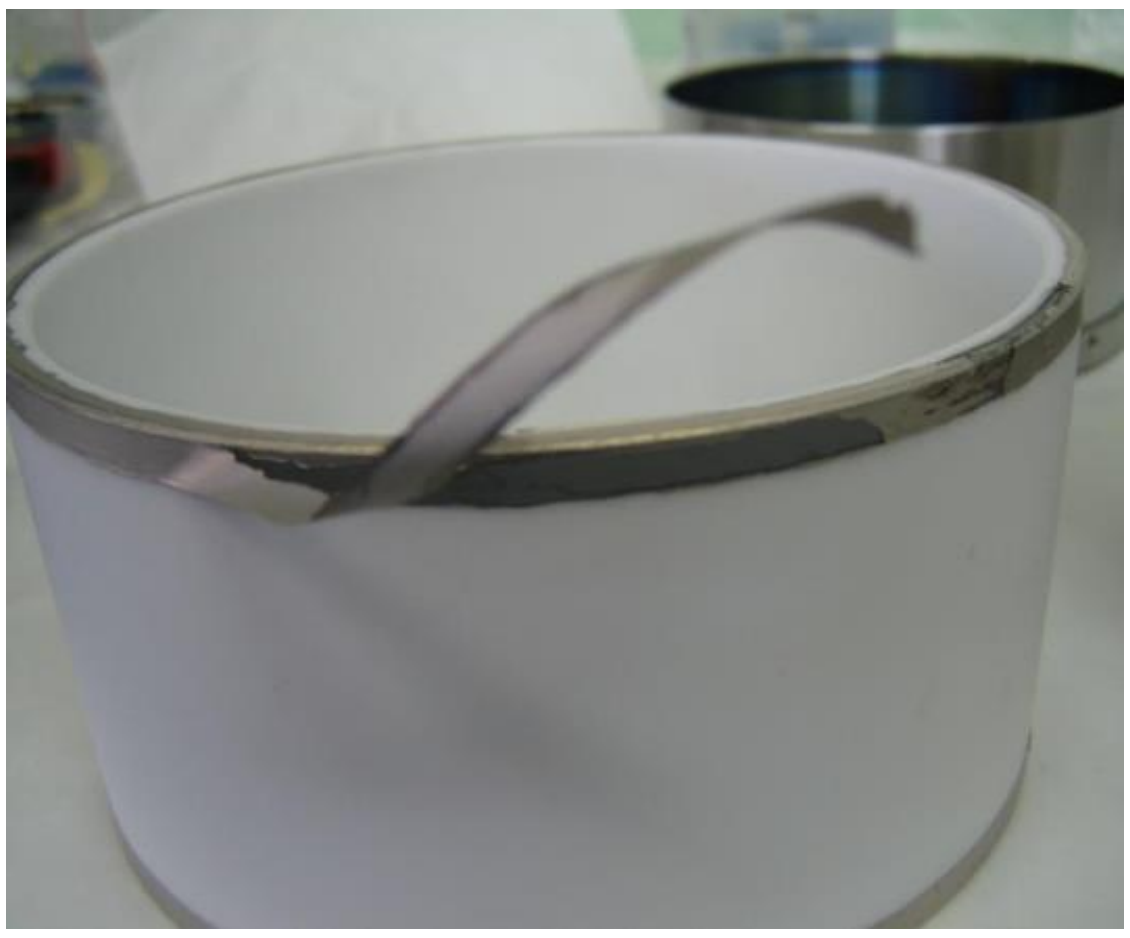

**Fig. 30:** Example of a default in the metallization of the ceramic, one can easily understand that it will not be possible to braze any metallic part onto it.

### 7.4 Other items of the FPC

FPC is a large topic that deserves an entire document to address it. A 146 slides presentation has been given by the author to the SRF 2021 tutorial session (indico.frib.msu.edu/event/38/attachments/159/1270/20210624_SRF_tutorial_FPC_Eric_Montesinos.pdf). Also, more details can be found in previous CAS listed in the References.

## Acknowledgements

The author would like to thank Hans-Peter Kindermann and Erk Jensen, his former bosses, for the invaluable constant support and advice all along the last 30 years having the honour to work with them.